# Possibility of Realizing Quantum Spin Hall Effect at Room Temperature in Stanene/Al$_2$O$_3$(0001)

Hui Wang[1,2], S. T. Pi[1], J. Kim[1], Z. Wang[2], H. H. Fu[1] and R. Q. Wu[1,2]

*[1]Department of Physics and Astronomy, University of California, Irvine, CA 92697-4575, USA*

*[2]State Key Laboratory of Surface Physics and Department of Physics, Fudan University, Shanghai 200433, China.*

Two-dimensional quantum spin Hall (QSH) insulators with reasonably wide band gaps are imperative for the development of various innovative technologies. Through systematic density functional calculations and tight-binding simulations, we found that stanene on α-alumina surface may possess a sizeable topologically nontrivial band gap (~0.25 eV) at the $\Gamma$ point. Furthermore, stanene is atomically bonded to but electronically decoupled from the substrate, providing high structural stability and isolated QSH states to a large extent. The underlying physical mechanism is rather general, and this finding may lead to the opening of a new vista for the exploration of QSH insulators for room temperature device applications.

Quantum spin Hall (QSH) state of two-dimensional topological insulators (2DTIs) is one of the most interesting research topics of condense matter physics and materials science [1-6]. The helical edge states of nanoribbons made from 2DTIs provide massless relativistic carriers with intrinsic spin-momentum lock and negligible backscattering [7] and, furthermore, they are also robust against edge modification, ideal for various applications that require dissipationless spin transport [3, 5, 8, 9]. Although many 2DTIs have been predicted theoretically [10-13], starting from graphene [1], very few experimental observations of QSH have been reported [3]. This is mainly caused either by the weakness of spin-orbit coupling (SOC) of light elements and henceforth their small topological band gaps (e.g., graphene or silicene), or by the structural instability of heavy elements in the metastable planar honeycomb structure (e.g., germanene [14] and Bi halide systems [15]).

Among potential 2DTIs for practical utilizations, a tin monolayer (stanene) has received particular attention. Because of its intrinsically strong SOC, stanene was predicted to have a large topological band gap (~0.3 eV) and reasonable structural stability after appropriate chemical functionalization [16]. Many novel physical properties such as thermoelectric performance [17], near RT quantum anomalous Hall effect (QAHE) [18] and topological superconductivity [19] were also found for stanene through density functional theory (DFT) calculations. Recent successful fabrication of stanene on the $Bi_2Te_3$(111) substrate using molecular beam epitaxy (MBE) further stimulated research interest in stanene physics [20]. Needless to say, a major drawback of stanene as a 2DTI is that its electronic properties are very sensitive to the change of environment. For example, stanene/$Bi_2Te_3$(111) was found to be metallic, even though stanene and $Bi_2Te_3$(111) interact through weak van der Walls forces and the QSH state of a free standing stanene is expected to be reserved [20]. Apparently, it is essential to search for suitable substrates for the realization of the QSH state of stanene [21-23].

In this Letter, we report results of systematic DFT calculations and tight binding (TB) modeling for the topological features of stanene on $α$-$Al_2O_3$(0001). Remarkably, the binding energy between stanene and the $α$-$Al_2O_3$(0001) substrate is as high as 0.557 eV per Sn atom, which is beneficial for growing stable stanene layer on this conventional substrate. The $α$-$Al_2O_3$(0001) surface induces a noticeable charge transfer between Sn atoms in the A- and B-sublattices and hence breaks the A-B symmetry of the Sn-lattice. Nevertheless, stanene/$Al_2O_3$(0001) possesses a band gap at the $Γ$ point and, importantly, this gap is topologically nontrivial according to examinations of the $Z_2$ number and the robust edge states in a stanene nanoribbon on $Al_2O_3$(0001). Our findings not only suggest a new material system for the realization of the QSH state in ambient condition, but also reveal novel physics for studies of 2DTIs in a honeycomb lattice.

Our DFT calculations were performed with the Vienna Ab-initio Simulation Package (VASP) [24] at the level of spin-polarized generalized gradient approximation (GGA) [25]. We treated O-$2s2p$, Al-$3s3p$ and Sn-$5s5p$ as valence states and adopted the projector-augmented wave (PAW) pseudopotentials to represent the effect of their ionic cores [26, 27]. Spin-orbit coupling term was incorporated self-consistently using the non-collinear mode [28, 29]. To obtain reliable adsorption geometries and binding energies, the optimized non local van der Waals functionals (optB86b-vdW) were also included [30-32]. The energy cutoff for the plane-wave expansion was set to as high as 600 eV, sufficient for this system according to our test calculations.

Stanene is a layer of tin atoms in the buckled honeycomb lattice ($a = b = 4.67$ Å), as shown in Fig. 1. The band structure of the pristine stanene has Dirac cones at the $K$ ($K'$) points and a nontrivial band gap (~0.1 eV) opens by the inclusion of SOC [16]. Stanene is chemically very active and its physical properties can be easily modified by the change of environment, e.g., the presence of adsorbates and substrates [20, 21]. These characteristics dictate the choice of appropriate substrates, i.e., they should have a hexagonal honeycomb structure on surface, small lattice mismatch with stanene, and sizeable bulk band gap. Sapphire $\alpha$-$Al_2O_3$ (0001) surface is one of the possible substrates that satisfy all these requirements. Unlike other substrates with uncertainties regarding surface termination and contamination, the $\alpha$-$Al_2O_3$ (0001) surface has an Al-termination and is chemically very stable [33-35]. The optimized lateral lattice constant of bulk $\alpha$-$Al_2O_3$ is about 4.81 Å [33], providing an excellent match with stanene. In the present work, we simulated the $\alpha$-$Al_2O_3$ (0001) surface with a slab model that consists of 18 atomic layers and a vacuum gap about 15 Å thick. A 9×9×1 Monkhorst-Pack k-point mesh was used to sample the Brillouin zone [36]. Positions of all atoms except that in the central six $Al_2O_3$ layers were optimized with a criterion that the atomic force on each atom becomes weaker than 0.01 eV/Å and the energy convergence is better than $10^{-6}$ eV.

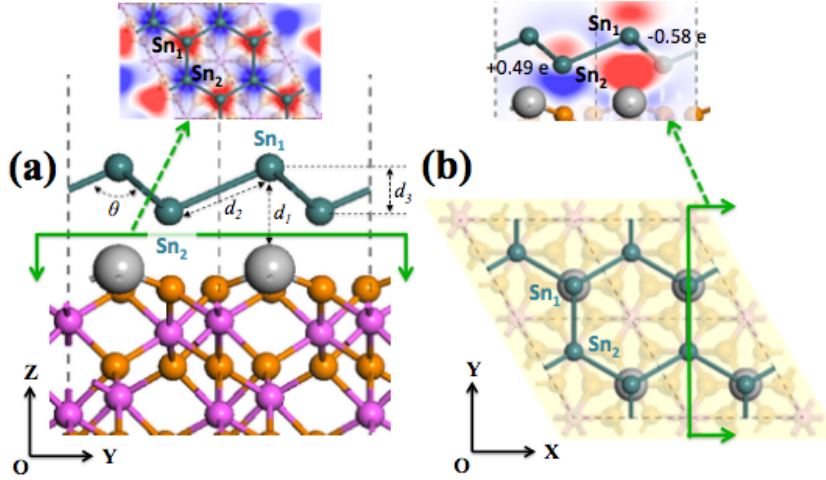

FIG. 1 (color online) Schematic model for (a) Side view and (b) top view of stanene adsorbed on α-alumina surface, cyan, orange and pink balls represent Sn, O and lattice Al atoms, respectively, surface Al atoms are depicted by large grey balls, coordinate axes are guided for your eyes. Insets demonstrate the charge redistribution of tin layer after adsorption at horizontal plane (green arrow in (a)) and vertical plane (green arrow in (b)), red and blue represents charge accumulation and depletion in a range of ±0.05 e/Å³. Positive and negative numbers in inset of (b) indicate the Bader charges of Sn atoms.

To describe the strength of stanene adsorption on the α-Al$_2$O$_3$ (0001) surface, we define the binding energy per Sn atoms as:

$$E_b = (E_{stanene/Al_2O_3\ (0001)} - E_{Al_2O_3\ (0001)} - E_{stanene})/N_{Sn} \qquad (1)$$

where $E_{stanene/Al_2O_3\ (0001)}$ and $E_{Al_2O_3\ (0001)}$ are the total energies of the Al$_2$O$_3$ slab with and without the stanene on it; $E_{stanene}$ is the total energy of the free standing stanene; $N_{Sn}$ is the total number of Sn atoms in the unit cell. We tested various initial adsorption configurations by placing Sn$_1$, Sn$_2$, hollow and bridge sites of stanene on top of the surface Al atom (denoted as "*Al-Sn$_1$*", "*Al-Sn$_2$*", "*Al-h*" and "*Al-b*" geometries below for the easiness of discussions). The most preferential structure is the "*Al-Sn$_1$*" configuration (as shown in Fig. 1), with a binding energy of -0.557 eV per Sn atom as shown in Table 1. The second best adsorption geometry "$Al\text{-}Sn_2$" has a much smaller binding energy, -0.262 eV per Sn atom; and the initial "$Al\text{-}h$" and "$Al\text{-}b$" setups converge to the "$Al\text{-}Sn_1$" geometry after the structural optimization. We perceive that stanene/$Al_2O_3$(0001) most likely adopts the "$Al\text{-}Sn_1$" geometry and therefore we focus our following discussions on results for this geometry without further notes.

Interestingly, stanene buckles more on $Al_2O_3$ even under the influence of a tensile stress, as indicated by the noticeable increase (decrease) of $d_3$ ($\theta$) in Table 1. As a result, the Sn-Sn bond length stretches to 3.034 Å on $Al_2O_3$ from 2.830 Å in the pristine stanene, in consistent with previous observations for stanene on other substrates [20, 21]. The distance between the topmost Al and $Sn_1$ atoms is 2.879 Å, and this Al atom is pull up by about 0.562 Å from its position in the clean α-$Al_2O_3$(0001) surface, which is nevertheless still 0.281 Å lower than its bulk-like position. It's worth noting that the $Al_2O_3$(0001) substrate induces strong charge transfer from $Sn_2$ to $Sn_1$ atoms, as shown in the insets of Fig. 1(a) and 1(b). Bader charge analysis demonstrates that each $Sn_1$ atom gains 0.58 electrons whereas each $Sn_2$ atom losses 0.49 electrons. Charge polarization can also be observed above the topmost Al and O atoms in the substrate. Along with the large binding energy (-1.114 eV per cell), it appears that stanene can form highly stable structure on the *α*-$Al_2O_3$(0001) surface. This can be an important advantage of using $Al_2O_3$(0001) rather than $Bi_2Te_3$(111) to support the growth of stanene, and we hope our results will inspire experimental interest in these systems.

TABLE 1. The binding energy ($E_b$), distance between Al and atop-Sn ($d_1$), bond length ($d_2$) and measures of buckling ($d_3$, $\theta$) for stanene adsorbed at different positions on the α-alumina surface and pristine stanene.

*Parameters $d_1$, $d_2$, $d_3$ and $\theta$ are also depicted in Fig. 1(a). Experimental data shown in parentheses are from the (111) surface of face-centered cubic tin.*

| Property | $E_b$ (eV) | $d_1$ (Å) | $d_2$ (Å) | $d_3$ (Å) | $\theta$ (°) |
| --- | --- | --- | --- | --- | --- |
| *Al-Sn$_1$* | -0.557 | 2.879 | 3.034 | 1.221 | 105.1 |
| *Al-Sn$_2$* | -0.262 | 2.990 | 2.864 | 0.705 | 114.2 |
| *Stanene* | -- | -- | 2.830 (2.854) | 0.854 (0.936) | 111.3 (109.8) |

On the other hand, it is expected that the electronic properties of stanene are strongly altered under the influence of the substrate. From the atomic-orbital projected density of states (PDOS) in Fig. 2(a), one can easily see the drastic differences between PDOSs of Sn$_1$ and Sn$_2$ in stanene/Al$_2$O$_3$(0001) and in the pristine stanene, particularly for the p$_z$ orbitals. Note that one of the most crucial features of 2DTIs, the symmetry between the A- and B-sublattice, no longer exists in stanene/Al$_2$O$_3$(0001). A legitimate question now "is stanene still topologically nontrivial under the strong influence of the substrate?".

To provide answers for this question, we present the calculated electronic band structures of stanene/Al$_2$O$_3$(0001) along the high symmetry lines in the Brillouin zone in Fig. 2(b). From their wave functions, we can identify that most bands around the Fermi level are from Sn atoms, as depicted by the red curves. When SOC is excluded in our DFT calculations, stanene/Al$_2$O$_3$(0001) behaves like a poor metal with two bands touching the Fermi level in the vicinity around the $\Gamma$ point. Remarkably, a gap as large as 0.25 eV opens after SOC is turned on in DFT calculations as shown in the up inset of Fig. 2(b), indicative of the nontrivial topological feature of stanene/Al$_2$O$_3$(0001). In addition, SOC produces band splittings of 0.05 eV at the valence band maximum (VBM) and 0.003 eV at the conduction band minimum (CBM) near the $\Gamma$ point, due to the Rashba term in low-symmetry structures and the hybridization between $p_z$ and $p_{x,y}$ orbitals in the buckled geometry [14, 37].

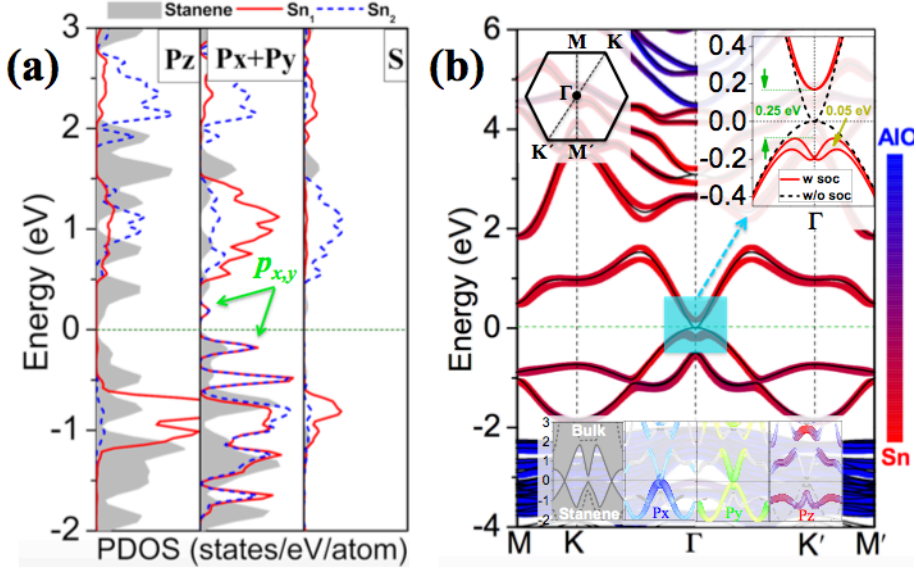

FIG. 2 (color online) (a) The atomic-orbital projected local density of states of pristine Stanene, $Sn_1$ and $Sn_2$ atoms are demonstrated for the $p_z$, $p_x+p_y$ and s orbitals, respectively. Green arrows highlight the $p_x$ and $p_y$ orbitals in the vicinity of the $\Gamma$ point. (b) Electronic bands of stanene/AlO with (thick colored lines) and without (thin black line) spin orbit coupling in calculations. The color bar indicates the weights of Sn (red) and substrate (blue). Green dashed line represents the Fermi level. Up left inset gives the Brillouin zone and high symmetry lines. Up right inset shows the zoom-in view of bands near the $\Gamma$ point and the topological band gap. Bottom inset gives the band structure of the pristine stanene, along with orbital resolved bands of stanene/AlO near the Fermi level.

The band topology can be characterized by the $Z_2$ invariant, with a nontrivial band topology represented by $Z_2 = 1$ while a trivial band topology represented by $Z_2 = 0$. To this end, we implemented the n-field method as a module for VASP to obtain the $Z_2$ invariants from the DFT Bloch functions [38-40]. This allows us to unambiguously confirm the topological feature of stanene/$Al_2O_3$(0001), as done in the literature for most topological materials [41]. The calculated n-field configuration for the buckled stanene on α-$Al_2O_3$ (0001) is shown in Fig. 3(a). One can easily obtain $Z_2 = 1$ by counting the positive and negative n-field numbers over half of the torus, clearly demonstrating the nontrivial band topology of stanene/$Al_2O_3$(0001). It should be noted that different gauge choices (e.g. plane wave cutoff, k-point mesh, etc.) result in different n-field configurations; however, the sum of the n-field numbers over half of the Brillouin zone is gauge invariant, namely $Z_2$ is robust against the choice of parameters and gauge.

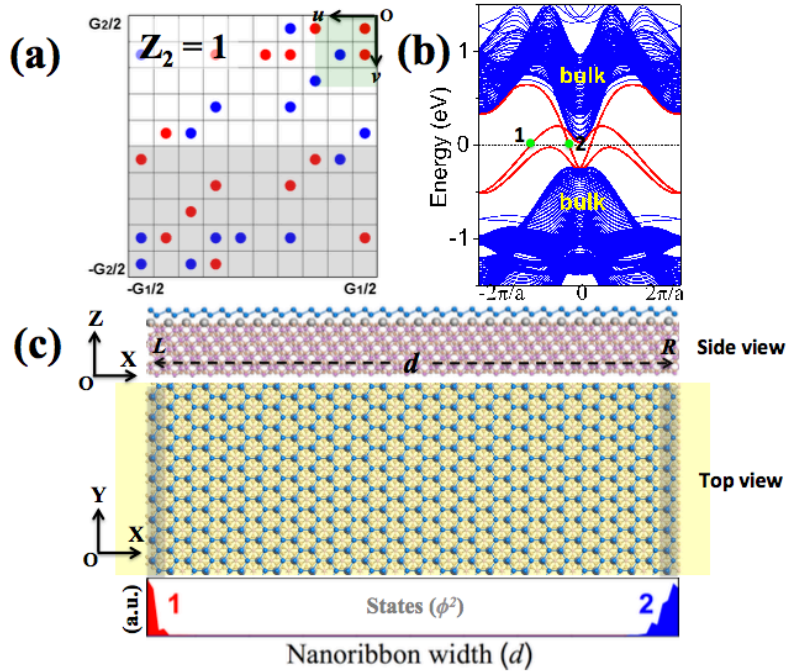

Fig. 3 (color online) (a) The n-field configuration of stanene α/AlO. The calculated torus in Brillouin zone is spanned by $G_1$ and $G_2$. Note that the two reciprocal lattice vectors u and v actually form an angle of 120°. The red and blue circles denote n = 1 and n = -1, respectively, while the blank denotes n=0. The $Z_2$ invariant is 1, by summing the n-field over half of the torus, as shown in the shaded yellow area. (b) Band structure of a zigzag nanoribbon of stanene/AlO demonstrates the existence of topological protected helical *edge states. (c) Top and side views for the schematic model of stanene/AlO nanoribbon along the y direction, color balls have the same meanings as in Fig. 1. Symbols "d", "L" and "R" in side view represent the length, left and right side of nanoribbon, respectively. The spatial localization of edge states "1" and "2" in (b) is demonstrated the projections of their wave functions.*

Another obvious characteristic of 2DTIs is the presence of gapless edge states with spin-momentum locking in the bulk topological band gap. Since the main features near the Fermi level are contributed by the *p* orbitals of Sn atoms (shown in the bottom insets in Fig. 2(b)), we use the maximally localized Wannier functions (MLWFs) to interpolate these bands [42] and reconstructed the tight-binding Hamiltonian of zigzag nanoribbons of stanene/$Al_2O_3$ with a width of $d \approx 21$ nm (25 unit cells) as sketched in Fig. 3(c). Although the renormalization on the onsite energy of edge states is not considered, it should not affect the key physical features of the topologically protected edge states. As shown in Fig. 3(b), the calculated electronic band structures of zigzag nanoribbons clearly demonstrate that several nontrivial gapless bands (highlighted by red) exist within the bulk gap (depicted by blue), connecting bulk valence bands with bulk conduction bands. These edge states have opposite group velocities for opposite edges, in agreement with previous reports [21, 43]. As an example, we obtained the real space projection of two specific eigenvalues at the Fermi level (green dots marked by "1" and "2" in Fig. 3(b)) by calculating the square of wave function along the *x* direction. As demonstrated in Fig. 3(c), one can easily see that these electronic states, "1" and "2", are mainly localized at different edges, propagating to different direction ascribed to the opposite Fermi velocity. The existence of gapless helical edge states together with the analysis of $Z_2$ topological invariant consistently proves that the supported stanene on $\alpha$-$Al_2O_3$ (0001) is a QSH insulator.

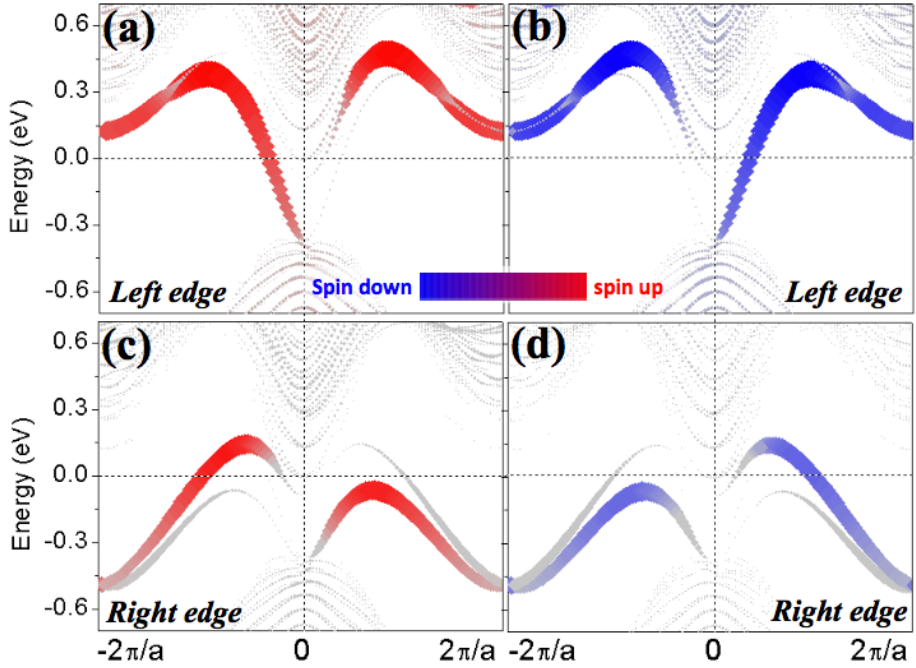

*Fig. 4 (color online) Spin projections of helical edge states of stanene$\alpha$/AlO. Red and blue represent spin up and spin down, respectively. The upper panels (a), (b) and lower panels (c), (d) demonstrate the left and right edges corresponding to the sketched atomic models in Fig. 3(c).*

As a further step, we made spatial and spin projections of these edge states and the results are shown in Fig. 4. One can obviously see the spin-momentum locking feature of 1D helical electrons (left edge) and holes (right edge) in stanene nanoribbon on $\alpha$-$Al_2O_3$(0001), the characteristics of QSH. Therefore, regardless the termination of the ribbon ($Sn_1$ or $Sn_2$), the QSH state of stanene/$Al_2O_3$(0001) should observable.

In summary, we have shown that stanene in a buckled honeycomb lattice develops a new topological QSH state on α-Al$_2$O$_3$(0001) substrate, confirmed by the direct calculations of the $Z_2$ topological invariant and gapless edge states. In addition, the considerable coupling strength between stanene and the α-Al$_2$O$_3$ (0001) substrate is helpful for experimental growth of stanene. The large topologically nontrivial band gap up to ~0.25 eV even without the symmetry between A-B sublattice is rather unusual and deserves more fundamental explorations. It is also foreseeable that stanene on α-Al$_2$O$_3$ holds a potential for room temperature QSH applications.

Work at UCI was supported by DOE-BES (Grant No: DE-FG02-05ER46237 for HW and HF; SC0012670 for JK, SP, and HF). Work at Fudan (HW and ZW) was supported by the CNSF (Grant No: 11474056) and NBRPC (Grant No: 2015CB921400). Computer simulations were partially supported by NERSC.